\documentclass[12pt]{iopart}

\usepackage{graphicx}
\usepackage{iopams}
\expandafter\let\csname equation*\endcsname\relax
\expandafter\let\csname endequation*\endcsname\relax
\usepackage{amsmath,amsfonts,amsthm,bm,mathtools}
\usepackage{latexsym}
\usepackage{times}

\usepackage[x11names]{xcolor}
\definecolor{linkcolor}{rgb}{0.3,0.3,1.0} 
\usepackage[pdftex,colorlinks=true, linkcolor= linkcolor, citecolor= linkcolor, urlcolor= linkcolor, hyperindex=true,hyperfigures=true]{hyperref} 

\renewcommand{\vec}{\bm}

\newcommand{\UNIPD}{
Department of Physics and Astronomy, University of Padova, Via Marzolo 8, I-35131 Padova, Italy}
\newcommand{\INFN}{INFN, Sezione di Padova, Via Marzolo 8, I-35131 Padova, Italy}
\newcommand{\DFG}{Signalling Research Centers BIOSS and CIBSS, University of Freiburg, 79104 Freiburg, Germany}
\newcommand{\UF}{Institute of Physical Chemistry, University of Freiburg, 79104 Freiburg, Germany}
\newcommand{\SGBM}{Spemann Graduate School of Biology and Medicine (SGBM), University of Freiburg}

\date{\today}

\begin{document}

\title{Constrained hidden Markov models reveal further Hsp90 protein states}

\author{Riccardo Tancredi,$^1$ Antonio Feltrin,$^1$ Giosu\`e Sardo Infirri,$^1$ Simone Toso,$^1$ Leonie Vollmar,$^{2,3}$ Thorsten Hugel,$^{2,4}$ Marco Baiesi$^{1,5,*}$}
\vspace{3 mm}
\address{$^1$ \UNIPD}
\address{$^2$ \UF}
\address{$^3$ \SGBM}
\address{$^4$ \DFG} 
\address{$^5$ \INFN}
\ead{$^*$ marco.baiesi@unipd.it}

\begin{abstract}
Time series of conformational dynamics in proteins are usually evaluated with hidden Markov models (HMMs). This approach works well if the number of states and their connectivity is known. However, for the multi-domain protein Hsp90, a standard HMM analysis with optimization of the BIC (Bayesian information criterion) cannot explain long-lived states well. Therefore, here we employ constrained hidden Markov models, which neglect transitions between states by including assumptions. Gradually tuning a model with justified and focused changes allows us to improve its effectiveness and the score of the BIC. This became possible by analyzing time traces with several thousand observable transitions and, therefore, superb statistics. In this scheme, we also monitor the residences in the states reconstructed by the model, aiming to find exponentially distributed dwell times. We show how introducing new states can achieve these statistics but also point out limitations, e.g., for substantial similarity of two states connected to a common neighbor. One of the states displays the lowest free energy and could be the idle open ``waiting state'', in which Hsp90 waits for the binding of nucleotides, cochaperones, or clients.
\end{abstract}


\section{Introduction}
The heat shock protein Hsp90 is one of the most abundant proteins in the cytoplasm of cells and a molecular chaperone fulfilling its function of assisting other proteins (called clients) to attain their active form. By interacting with a wide range of cochaperones and client proteins, it maintains proteostasis and thus essential cellular functions~\cite{Graf2009, Schopf2017, Borkovich1989}. Hsp90 is a homodimer, each monomer consisting of three domains. The abundant structural data on Hsp90 distinguishes two conformational states: an N-terminally open configuration and an N-terminally closed configuration~\cite{Shiau.2006, Ali2006}. Transitions between these conformational states are part of Hsp90’s catalytic cycle, e.g., the N-terminally open conformation allows for the binding of many clients, some clients are then threaded through the dimer in the closed state \cite{verbaAtomicStructureHsp90Cdc37Cdk42016,garcia-alonsoStructureRAF1HSP90CDC37Complex2022}. In general, the function of proteins is determined by the different structures they adopt and how they interconvert between them~\cite{Xie2020,Fraser2009,Knoverek2019}.

For Hsp90, several fluorescence-based studies have shown that at least four states (two N-terminal open and two N-terminal closed) are necessary to describe the data~\cite{Mickler2009, schmid2016single}. Therefore, a transition between these four states is the current picture of the idle Hsp90 dimer, but more states could also describe these previous data. The four states have either been defined by the structural analogy between different Hsp90 homologues~\cite{Shiau.2006} or by a hidden Markov model (HMM) analysis of single-molecule FRET data~\cite{schmid2016single, Vollmar2024} or Plasmon ruler data~\cite{ye2018conformational}.

The HMM is a flexible tool that aims to infer undetected dynamics from partially available information~\cite{dymarski2011hidden,bouguila2022hidden,gotz2022blind}. It has been used for a variety of purposes, such as studying stochastic systems with feedback and measurement errors~\cite{bechhoefer2015hidden} or fluorescent time series of quantum dots~\cite{martinez2020improving},
improving qubit readout~\cite{martinez2020improving}, and searching for gravitational waves~\cite{abbott2017search,suvorova2016hidden}.
Moreover, HMMs are useful in biophysics for improving our understanding of molecular motors~\cite{mullner2010improved,syed2010improved} and heterogeneous diffusion states in the cytoplasm~\cite{janczura2021identifying}. 

\begin{figure*}[t!]
    \centering
\includegraphics[width=0.99\textwidth]{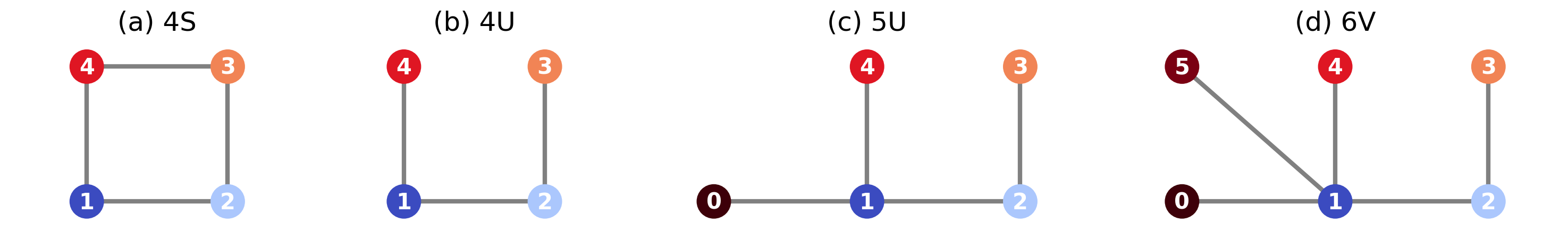}
    \caption{(a) Diagram of the network of states and transitions between them, as previously found for Hsp90 in solution with ATP~\cite{schmid2016single}. We use the label 4S for {\em square}, as the network contains a closed loop. The colored circles represent the different hidden states of the HMM, and their position (up or down) indicates the (high or low) scattering intensity. Links represent the transitions between the states found by an HMM.
    (b), (c), and (d): Diagrams of the CHMMs implemented in this work. ``U'' in 4U and 5U emphasizes the u-shaped part of the network; ``V'' in 6V draws attention to the relevant v-shaped 4-1-5 motif. Along a row, darker colors correspond to lower escape rates. In these models, an absence of a link indicates the enforced presence of a zero in the corresponding entry of the transition matrix.} 
    \label{fig:net}
\end{figure*}

\begin{figure*}[ht!]
    \centering
    \includegraphics[width=0.99\textwidth]{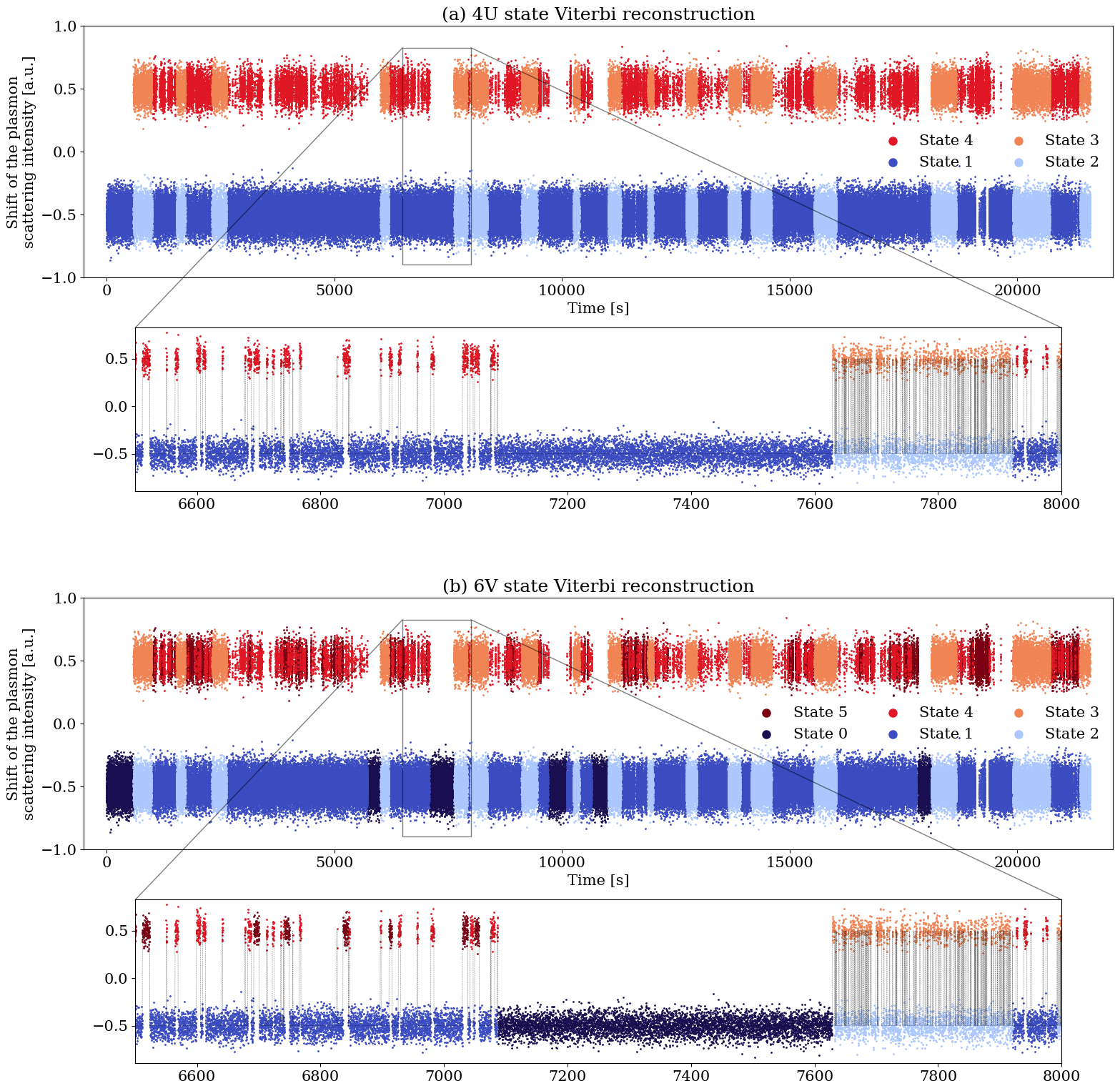}
    \caption{(a) Experimental trace colored according to the reconstructed sequence of states (i.e., the Viterbi path of the CHMM) for model 4U and (b) for model 6V. The lower panel is a zoomed region that shows how the 6-state model uses state 0 to explain a very long low-intensity period.} 
    \label{fig:vit}
\end{figure*}

Here, the analysis of a very long single-molecule time series from Plasmon ruler spectroscopy allows us to
revise the number of hidden states for the idle Hsp90 using an unconventional HMM approach.
The base is a standard HMM for physical systems hopping between states, each emitting a continuous signal. Formerly, four states emerged from the Bayesian information criterion (BIC) as the best tradeoff between model complexity and sound reproduction of the dynamics~\cite{schmid2016single,ye2018conformational}. The corresponding topology of the system's states is sketched in figure~\ref{fig:net}(a) (model 4S). Lines represent transitions with rates significantly different from zero. The corresponding reconstructed signal was similar to the one shown in figure~\ref{fig:vit}(a). It is composed of stretches with  $1\leftrightarrow 4$ jumps and others with $2\leftrightarrow 3$ jumps, with sporadic transiting between them through the $1\leftrightarrow 2$ and $3\leftrightarrow 4$ channels. A closed loop in the network of states was needed to model the Hsp90's nonequilibrium cycle consuming ATP.

Contrary to previous studies, we use the constrained HMM (CHMM) method~\cite{roweis1999constrained} to assume that transitions between certain states are impossible. This procedure reduces the number of free parameters in the model, which can modify and possibly improve the BIC score for a given number of states. Our strategy with CHMMs is building upon a good model from an HMM by gradually modifying the most sensible elements that improve the (BIC score of the) model while keeping it simple enough. 
Since we analyze an experiment in equilibrium (no ATP), we can consider a network of states without loops. From results with standard HMMs, we find that transitions $3\leftrightarrow 4$ take place with the lowest rates in model 4S.
Hence, our first natural simplification assumes transitions $3\leftrightarrow 4$ are so rare that they can be neglected in equilibrium. This leads to the 4U model of figure~\ref{fig:net}(b), which can better represent the data than the 4S model when comparing their BIC scores.
Figure~\ref{fig:vit}(a) is precisely the reconstructed time series of such a CHMM. The following shows subsequent upgrades to more complex models with five or six states, which explain the data better.

\section{Data}
\label{sec:Data}
The data analyzed here have been recorded by Plasmon ruler spectroscopy in the Soennichsen lab before, where their acquisition is described in detail~\cite{ye2018conformational}.  In short, the eponymous plasmon ruler in plasmon ruler spectroscopy consists of two plasmonic nanoparticles (in this case, gold) bridged by the macromolecule under investigation (here, the Hsp90 dimer). Due to plasmon coupling between the two gold nanospheres, their plasmon resonance is shifted to higher wavelengths, and the scattering intensity is increased when the interparticle distance is decreased by a closing of the Hsp90 dimer. For these measurements, yeast Hsp90 was mutated for a cysteine at position 285. A gold nanosphere (diameter of about 60 nm) was attached to each of the two Hsp90 monomers at these cysteines via flexible PEG linkers. These constructs were then immobilized on a glass substrate and observed in a dark field microscope. The scattering intensity of the gold nanospheres was recorded by a CMOS camera. This procedure allowed for single-molecule observation of Hsp90's N-terminal open-close dynamics with 10~Hz for six hours. Here, we analyze traces from Hsp90 in the absence of nucleotides from reference \cite{ye2018conformational}. One time series of the opening and closing dynamics of an Hsp90 dimer, recorded by Plasmon ruler spectroscopy and termed ``Thermal1'', is shown in figure~\ref{fig:vit}(a). The y-axis depicts the shift of the plasmon scattering intensity in arbitrary units. The open Hsp90 dimer shows a low signal (here depicted in shades of blue), while the closed dimer gives a high signal (here depicted in shades of red).
\ref{app:A} summarizes some findings from another time series (``Thermal2'') recorded by Plasmon ruler spectroscopy.


\section{Constrained hidden Markov models}
HMMs have been routinely used for decades to determine the hidden states $z_t$ visited by a system from some available partial information in the form of a time series of visible states $x_t$, with $t=0,1,2,\ldots, T$ denoting a discrete time index. 
An HMM with $N$ hidden states $\{Z_1,\ldots, Z_N\} \ni z_t$ (here, we do not deal with continuous $Z$'s) assumes that the hidden jumping process $z_{t}\to z_{t+1}$ is a Markov chain with jumping probabilities $a_{ij} = p(Z_j|Z_i)$. It aims to optimize the likelihood of observed data $(x_1,x_2,\ldots, x_T)$ by tuning the parameters that determine the $N^2-N$ jumping probabilities (the $-N$ term comes from their normalization from each state) and the {\em emission} probabilities $p(x|Z_i)$. In our case, the emissions to a real continuous variable $x$ depend on $2 N$ parameters, namely a mean and standard deviation of a Gaussian distribution for each hidden state $Z_i$. Moreover, one adds $N-1$ parameters for the normalized probability of the initial states.
Hence, in total, each HMM at most contains a number of parameters
\begin{equation}
    R_{\max}(N) = N^2 + 2N - 1
\end{equation}
which influences the outcome of the BIC. Indeed, for an HMM with $R$ parameters and a maximized value of the expected log-likelihood function $\mathcal{L}$,  the minimization of the BIC score
\begin{equation}
    \textrm{BIC} = R \ln T - 2 \ln \mathcal{L}
\end{equation}
allows the selection of a model of moderate complexity that describes the data well.

In a constrained HMM (CHMM)~\cite{roweis1999constrained}, one uses $R<R_{\max}(N)$ parameters by introducing constraints based on assumptions. Specifically, the constraints assume that some transitions between states are impossible, i.e., they build some topology into the hidden state representation. For example, in figure~\ref{fig:net}(b), the three lines connecting $N=4$ states represent their possible transitions. For consistency with thermodynamics, these transitions are bidirectional. It yields $6$ independent transition probabilities (self-transitions derive from normalization), which is lower than the maximum number $N^2-N=12$ and leads to $R=17 < R_{\max}(4)=23$. Therefore, assuming a model with the structure of figure~\ref{fig:net}(b) may improve the BIC score with respect to the full HMM if the chosen transitions match the essential ones in the system.

CHMMs, analogous to HMMs, use the Baum-Welch forward-backward algorithm~\cite{bouguila2022hidden} to train and improve the model's parameters. Then, after the learning phase, they use the Viterbi algorithm~\cite{bouguila2022hidden} to retrieve the most probable sequence $(z_t)$ of hidden states. We implement both algorithms with the support of the log-sum-exp trick \cite{blanchard21} to prevent underflow. The Baum-Welch algorithm updates each entry $a_{ij}$ of the transition matrix by a term proportional to $a_{ij}$ itself. Therefore, if a specific entry $a_{ij}$ starts as $a_{ij} = 0$, it will remain zero at every update. The core idea of CHMMs is thus expressed in terms of the chosen initialization: If one sets an initial element in the transition matrix to zero, it will remain such during training and drop from the count of the parameters. We describe this point in detail in~\ref{app:B}.
In the end, each transition rate $w_{ij}$ for an allowed transition $i\to j$ is simply computed from the transition probability $a_{ij}$: $w_{ij} = a_{ij}/\Delta t$.

In the following section, we use CHMMs in conjunction with statistical considerations to show that the experimental time series for Hsp90 are compatible with $N=6$ states. 
Formerly, a $N=4$ state model emerged from analyzing single-molecule FRET data of Hsp90 not driven by ATP consumption (Apo state) using normal HMMs \cite{schmid2016single}. Now, analyzing much longer plasmon ruler time traces, we also see states with very long dwell times, which are rarely visited and had been missed due to the short single-molecule FRET traces. In the following section, we discuss our reasoning for expanding the system to the $N=5$ states of model 5U (figure~\ref{fig:net}(c)) and finally settling to $N=6$ as in model 6V (figure~\ref{fig:net}(d)).

\section{Data analysis}

The models that we discuss are sketched in figure~\ref{fig:net}(a)-(d). The schemes display states with a high plasmon resonance shift (high intensity) in the top row and states with a low plasmon resonance shift (low intensity) in the bottom row. In all cases, the emission parameters are essentially indistinguishable between states of high intensity and the same for states with low intensity. For example, $p(x|i=2)\approx p(x|i=3)$. This extreme overlap of the visible signal for different hidden states makes the reconstruction of the hidden sequence $z_t$ challenging.

Previous HMM analysis of single-molecule fluorescence time series for the protein Hsp90 converged to model 4S (figure~\ref{fig:net}(a))~\cite{schmid2016single}.
Model 4S emerged as the best one because the BIC applied to normal HMMs penalized and led to the discarding of more complex models with $N>4$. Thus, model 4S resulted in the optimal trade-off between simplicity and effectiveness.
However, as argued in the introduction, the simpler Model 4U in figure~\ref{fig:net}(b) is a better starting point for our CHMM analysis of Hsp90 data in equilibrium.

First, we determine the optimal parameters with a CHMM assuming the 4U structure in figure~\ref{fig:net}(b). We recover the two pairs of states: states 2 and 3, connected by fast jumping rates, and states 1 and 4, joined by a slower transition path. Furthermore, both pairs of transition rates (``paths'') are faster than the path between low-intensity states 1 and 2.

\begin{figure*}[t!]
 \centering
 \includegraphics[width=0.99\textwidth]{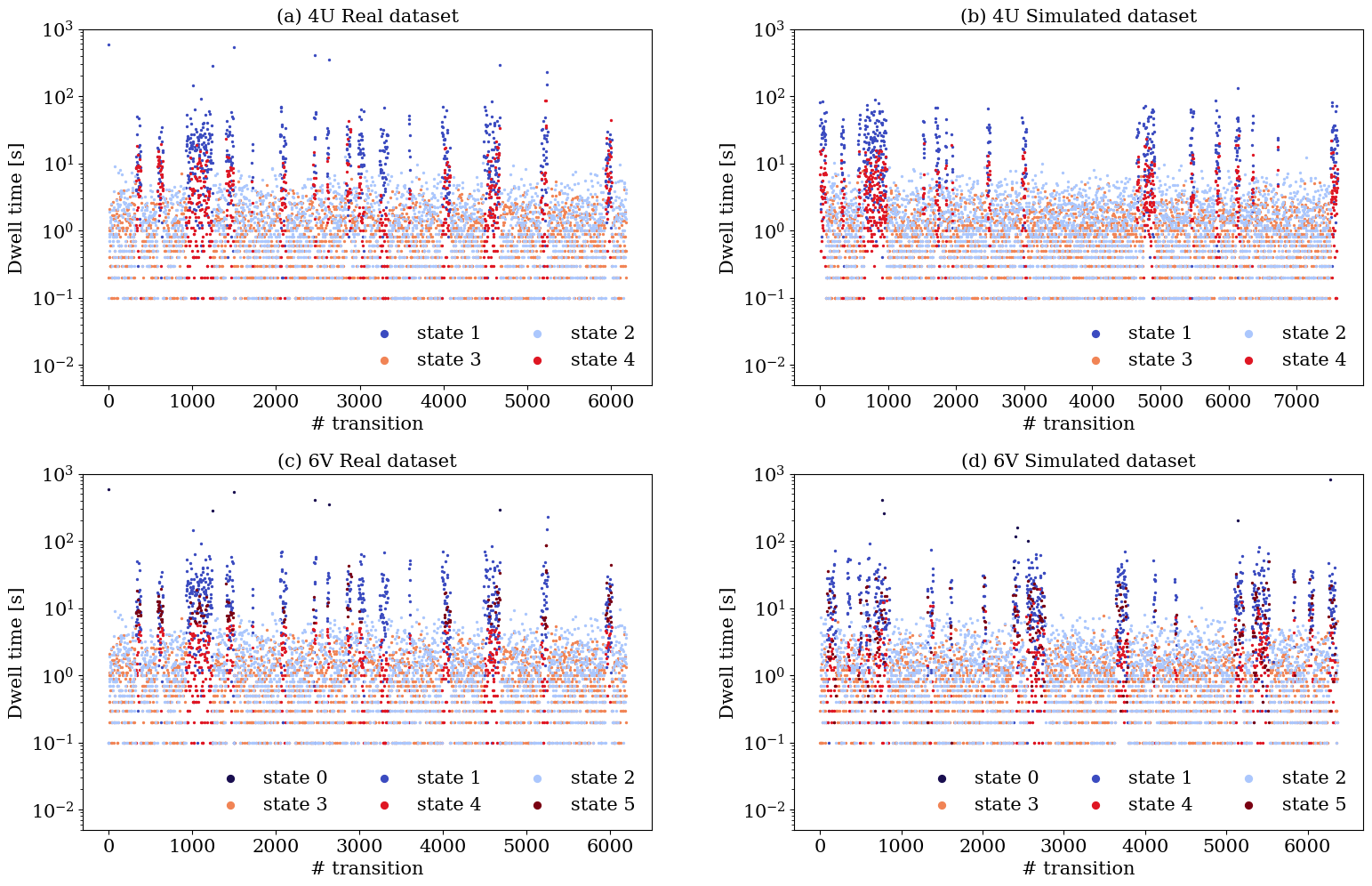}
 \caption{Sequences of dwell times: (a) reconstructed by the CHMM for the networks 4U shown in figure~\ref{fig:net}(a), and (b) from simulations based on the parameters for this model. Note the lack of simulated times $>10^2$s. (c) and (d) show the respective plots for model 6V in figure~\ref{fig:net}(d). }
 \label{fig:dt}
\end{figure*} 

\begin{figure*}[t!]
 \centering
 \includegraphics[width=0.99\textwidth]{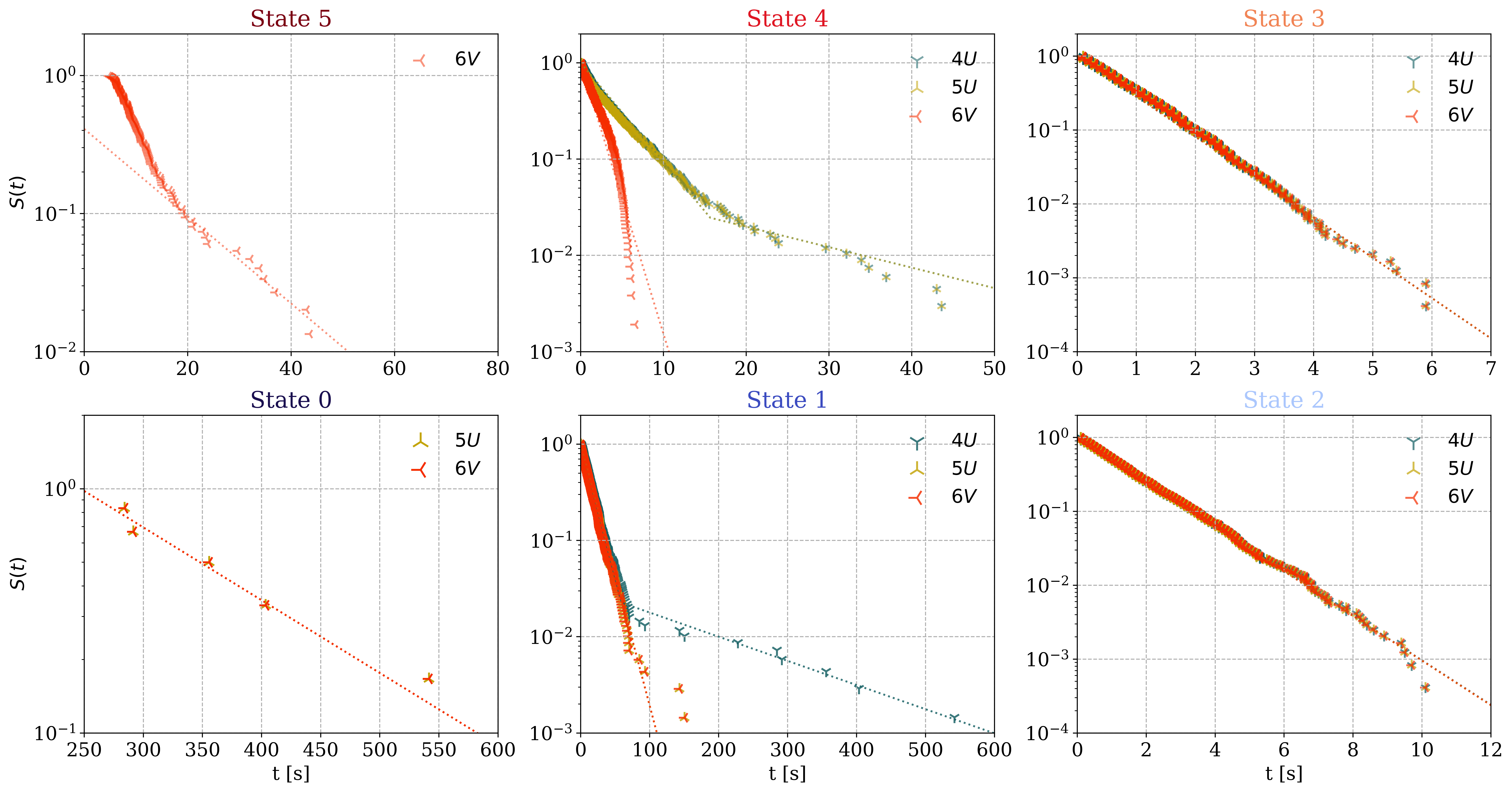}
 \caption{Each panel shows the statistics of the dwell times reconstructed by the models (see legends) for the state specified in its title: symbols are the survival probability $S(t)$, and straight lines are exponential fits. For the simpler models 4U and 5U, we show two fits when there are two exponential decays in a $S(t)$.
 Panels' layout follows the state's structure of model 6V in figure~\ref{fig:net}(d).
 }
 \label{fig:pdf}
\end{figure*} 

However, a ‘‘resimulation’’ of a system with the same parameters (as done, e.g., in~\cite{schmid2016single}) with a stochastic Markov chain reveals a limitation of the proposed model: it cannot reproduce the longest dwell times $\gtrsim 10^2$s of the experiments, as shown in figure~\ref{fig:dt}(b). In fact, the reconstructed distributions of dwell times $P(\tau)$, better visualized with the survival probability $S(t) = \int_t^{\infty} P(\tau)d\tau$, for model 4U shows (figure~\ref{fig:pdf}) that they are not single exponentials for every state.
In particular, distributions for states 1 and 4 (central panels of figure~\ref{fig:pdf}) contain two exponential regimes. 
This leads to the impossibility of simulating the longest times: the CHMM does not give enough significance to the sparse, extra-slow transitions. The simulations, therefore, can only reproduce the dwell times generated according to the first exponential.

In order to fix the problems with state 1, we assume it can be split into states 0 and 1 in a model that we term 5U, shown in figure~\ref{fig:net}(c). State 0 will represent a {\em long-term parking} state aiming to account for the series' longest dwell times. The training of the CHMM for model 5U converges to an expected log-likelihood $\mathcal{L}$ that improves the BIC score (BIC(5U)$-$BIC(4U)$=-142$) despite the number of parameters increased by five: two more for the transition probabilities between 0 and 1, two for the emission probabilities of state 0, and also one for the initial probability of state 0.
A version with state 0 attached to state 2 rather than 1 does not improve the results. Possibly, state 0 is attached to both other ones. Still, the few available transitions from data do not allow for validation of this hypothesis, which is discarded by the BIC due to the additional transition parameters.
Now, the dwell time distribution for state 1 scales correctly as a single exponential (lower central panel of figure~\ref{fig:pdf}). However, state 4 still manifests a double exponential distribution, prompting us to move to a $N=6$ model.

The natural extension from model 5U to $N=6$ states involves splitting state 4 into two states, now labeled 4 and 5. The name 6V for this model (sketched in figure~\ref{fig:net}(d)) emphasizes the ``V'' shape of the connections from state 1 to states 4 and 5, which is critical for understanding the following results. 
Model 6V is supported by the further decrease of the BIC score, BIC(6V)$-$BIC(4U)$=-177$. In the time series of the experimental signal colored according to the reconstructed states,  one may appreciate the difference between the predictions of the simpler model 4U (in figure~\ref{fig:vit}(a)) and model 6V (in figure~\ref{fig:vit}(b)). The latter assigns the long, low-intensity dwell times to parking state 0. Moreover, as visible in the lower panel of figure~\ref{fig:vit}(b), model 6V predicts sequences of jumps $1 \leftrightarrow 4$ mixed with those $1 \leftrightarrow 5$, corroborating the hypothesis that the network 6V structure with both 4 and 5 connected to state 1 is correct. In addition, simulations with model 6V parameters reproduce the $>10^2$s dwell times; see figure~\ref{fig:dt}(d) (also model 5U reproduced this feature, data not shown).

However, figure~\ref{fig:pdf} shows that the dwell times for states 4 and 5, even for model 6V, still do not follow a single exponential form.
The distribution for state 4 has an exponential statistics cut off at $\approx 6$s. Conversely, the distribution for state $5$ has no times lower than $\approx 6$s, an excess of statistics at intermediate times $\gtrsim 6$s, and asymptotic exponential statistics compatible with a long-lived behavior. For the rest, states 4 and 5 are part of the high-intensity signal and are indistinguishable due to their similar emission parameters.

The interpretation of these features is that the Viterbi algorithm, being unable to distinguish state 4 from 5 from the signal but trying to maximize the likelihood of the data, chooses the most probable event among $4\to 1$ and $5\to 1$ jumps mostly based on the (fitted) typical dwell time.
Thus, a short observed dwell time is paired to a $4\to 1$ event because the average dwell time of state 4 is shorter than state $5$. Conversely, a time longer than $6$s is assigned to $5\to 1$, even when the likelihood of the $4\to 1$ occurrence is barely lower.
Similarly, this mechanism also explains the lack of statistics below $250$s in the dwell times for state 0 (figure~\ref{fig:pdf}): the Viterbi algorithm finds it more logical to assign a short dwell time of the low-intensity signal to state 1 or 2; hence, it cannot detect short dwell times in state 0.

In Table~\ref{tab:escape_rates}, we summarize the escape rates $\lambda_i$ that the CHMM finds for model 6V. For all models and both data sets they are also reported on the diagonal of the tables in figure~\ref{fig:W} in~\ref{app:A}. In Table~\ref{tab:escape_rates}, we also report the index $\omega_i$ determined from an exponential fit $\sim e^{-\omega_i t}$ of the dwell time  survival probability $S(t)$ (the dashed lines in figure~\ref{fig:pdf}) for model 6V. For every state $i$, we find good agreement between $\omega_i$ and $\lambda_i$.

\begin{table}
\caption{Fitting parameters of the dwell times exponential behavior ($\omega_i$) for the states of the 6V model compared with escape rates $\lambda_i=\sum_{j\ne i} w_{ij}$ summing the transition rates found by the CHMM for the data set from the main text (``Thermal1'') for each state $i$. The fits yielding $\omega_i$'s are the lines for model 6V in figure~\ref{fig:pdf}; for state 5, the fit excludes times lower than $20$s, where the statistics deviates from the asymptotic exponential behaviour.}
\label{tab:escape_rates} 
\centering
\begin{tabular}{@{} lll @{}}
\br
State&$\omega_i$ [$s^{-1}$]&$\lambda_i$ [$s^{-1}$]\\
\mr
    0 & $0.007$  & $0.004$ \\
    1 & $0.063$  & $0.069$ \\
    2 & $0.69$   & $0.67$ \\
    3 & $1.26$   & $1.10$ \\
    4 & $0.65$   & $0.48$ \\
    5 & $0.07$   & $0.12$\\
\br
\end{tabular}
\end{table}

\begin{figure}[t!]
 \centering
 \includegraphics[width=0.45\textwidth]{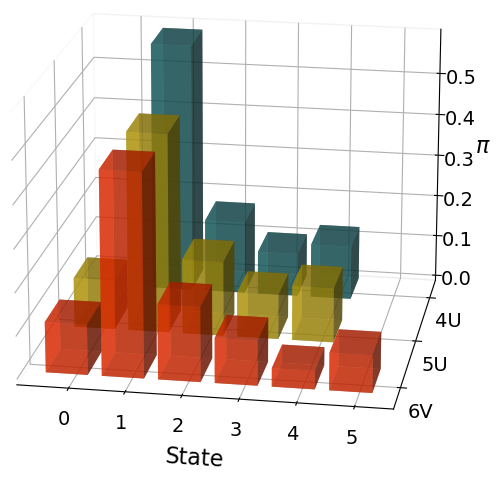} 
 \caption{Stationary state probability for the three different models investigated in detail.}
 \label{fig:pi}
\end{figure}

In all models, state 1 emerges as a special one: it is the state with the highest stationary probability, as reconstructed by the CHMM for models 4U, 5U, and 6V (see figure~\ref{fig:pi}). Therefore, it is the lowest free energy conformation of Hsp90. Moreover, state 1 is an open state and the most central one in the network of states for models 5U and 6V. Hence, we speculate that state 1 is a kind of ``waiting state'' of the idle Hsp90. I.e., it might be the state in which Hsp90 waits for the binding of nucleotides or cochaperones or clients to proceed towards other states \cite{Vollmar2024}.

We do not consider models with $N>6$ since there is no strong clue for more states to exist. Another version with $N=6$, in which state 5 is connected to state 0 rather than to state 1, does not perform as well as model 6V: the BIC score is worse, and the dwell times distribution for state 0 becomes non-exponential. Hence, model 6V emerges as the best fit for our data.

It should be interesting to understand which neglected transitions between the six states should be reintroduced in order to model Hsp90 in the presence of ATP. We postpone this to future work.

\section{Conclusions}
We have analyzed the dynamics of the protein Hsp90 measured by Plasmon rulers. The experimental Plasmon ruler time traces clearly show an N-terminal open and an N-terminal closed regime. A previous dynamic analysis based on HMM required at least four (hidden) states to describe the time traces \cite{ye2018conformational}, but more states could also explain the data. Here, we analyze this data with CHMMs, namely hidden Markov models with constraints, and we show that adding two more states in a well-defined network better describes Hsp90's opening and closing dynamics. This is important because not only the structure of a protein but also the number of states and their conversion rates determine the function of proteins~\cite{Fraser2024}.
A more complete knowledge of the number of states and their interconversion, as provided here, therefore improves our understanding of Hsp90 and should enhance our ability to target Hsp90 with drugs, for example.

For our analysis, it was crucial that the experimental technique allowed sampling over time scales of hours, covering the long-lived states' dwell times of about 10 minutes.
Our model selection was performed step by step, gradually improving the BIC score by incorporating minimal pondered additions supported by the observed statistics. It is a strategy not limited to protein data analysis; we expect it to provide good results for general systems.

The learning from data of CHMMs, of course, as for normal HMMs, needs a good convergence of the Baum-Welch algorithm. In principle, this convergence may suffer from the introduction of constraints, which limits the number of parameters and the space of possible learning dynamics. Nevertheless, this is not an issue in our tests and is well compensated by the improved BIC scores generated with fewer parameters.

The additional Hsp90 protein states are consistent with the experimental observation of N-terminal dynamics on a wide range of time scales from nanoseconds~\cite{Sohmen2023Onset} to several minutes, as shown here. These additional states might, for example, reflect different angles between the N-terminal and middle domain~\cite{Hellenkamp2017, Daturpalli2017} or asymmetric states~\cite{Mishra2014, Mayer2015}. Finally, CHMMs might be a helpful addition to the wealth of already available HMM algorithms available for the analysis of time traces~\cite{gotz2022blind}.

\paragraph{Acknowledgements} 
This work was supported by the Deutsche Forschungsgemeinschaft
(DFG) under Germany’s Excellence Strategy (CIBSS EXC-2189 Project ID 390939984) and the SFB1381 programme (Project ID 403222702). We thank the Soennichsen lab for providing the raw data of the Plasmon ruler traces.



\appendix

\section{Additional information and plots for the second dataset ``Thermal2''.}
\label{app:A}

\begin{figure*}[t!]
 \centering
 \includegraphics[width=0.99\textwidth]{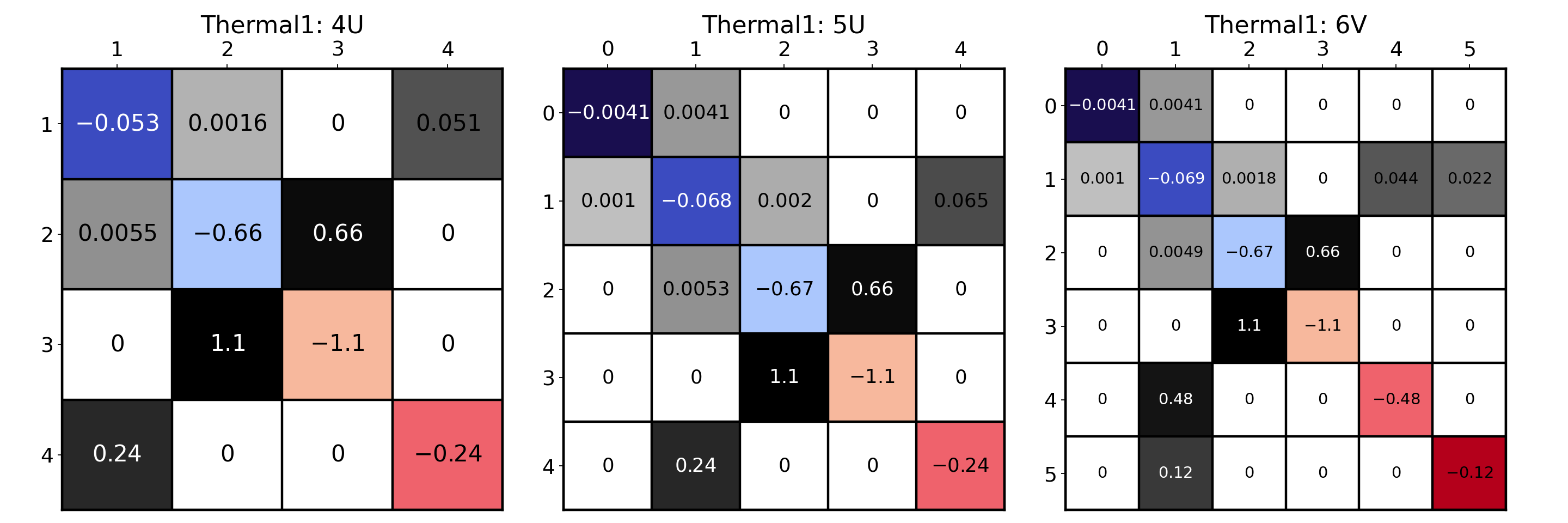}
 \label{fig:1ThermalW}
 \vskip 5mm
 \centering
 \includegraphics[width=0.99\textwidth]{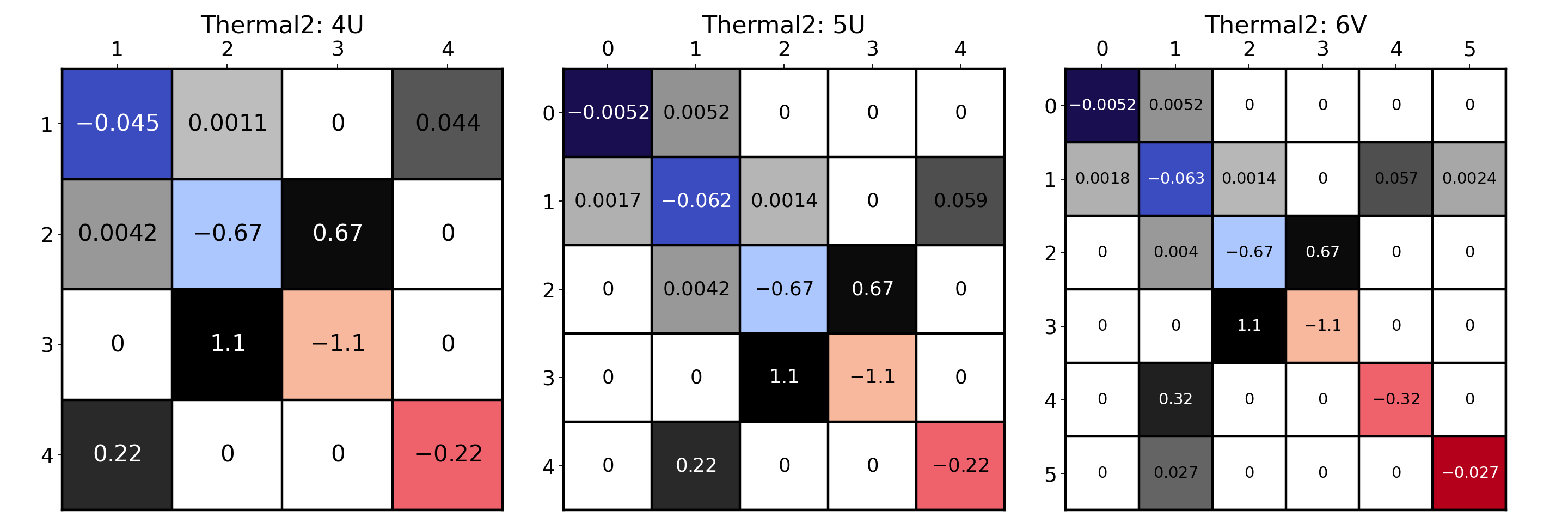}
 \caption{Transition rates in $s^{-1}$ for the three models, for the dataset used in the main text (``Thermal1'', top panels), and for the second dataset (``Thermal2'', bottom panels). Diagonal elements are the negative of states' escape rates $\lambda_i$, and their cell follows the standard color code for distinguishing the six states; off-diagonal terms are transition rates $w_{ij}$ from state $i$ (row) to state $j$, and their cell's darkness scales with the log of their value. Transitions not included in the CHMM have white cells with zero rates.}
 \label{fig:W}
\end{figure*}

\begin{figure*}[ht!]
    \centering
    \includegraphics[width=0.99\textwidth]{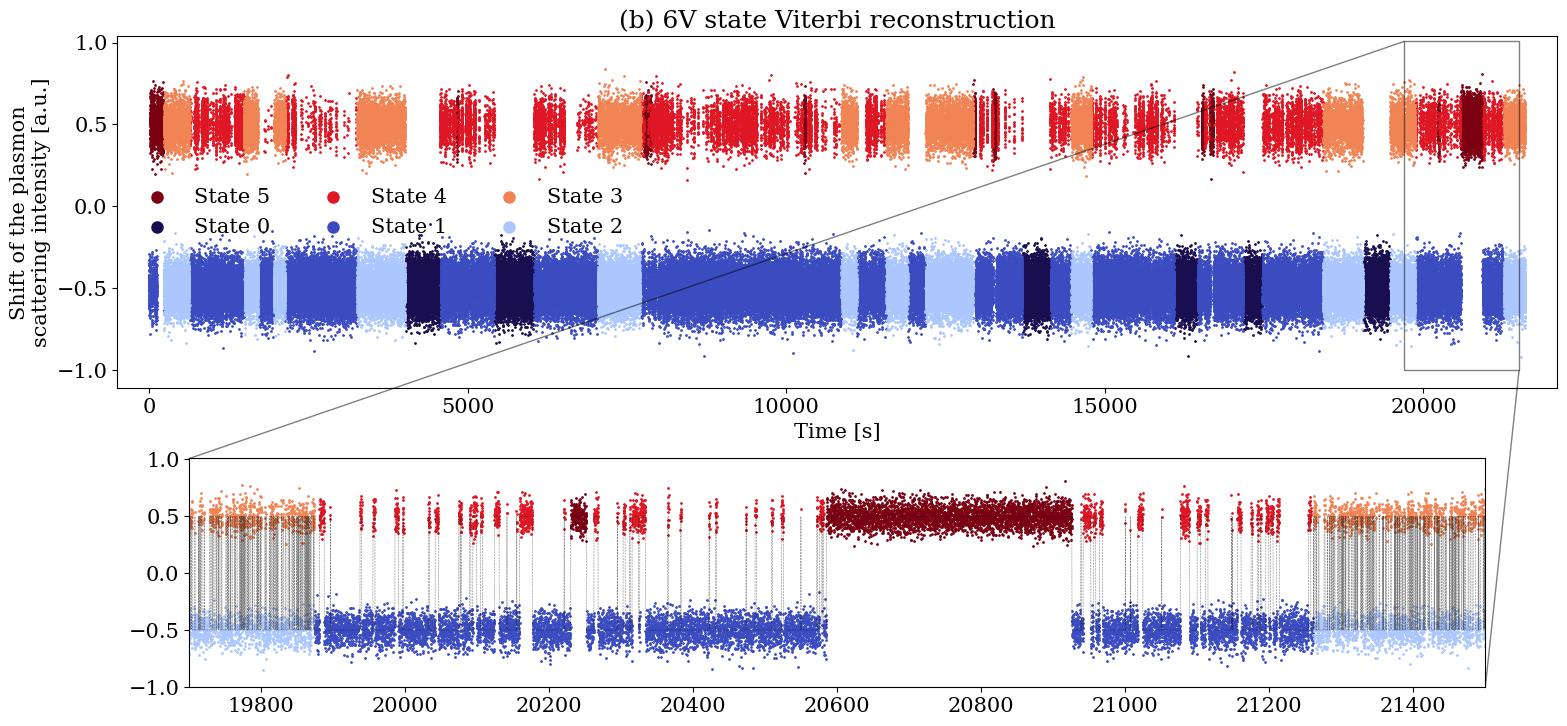}
    \caption{Viterbi reconstruction of the hidden states with model 6V for another experimental trace, which is provided as supplementary data file ``Thermal2''. The zoom highlights a long dwell time in the state 5.}
    \label{fig:6V2}
\end{figure*}

We briefly collect some information on the rates determined by the CHMM for all models (figure~\ref{fig:W}), and we show the Viterbi sequence for the second data set, ``Thermal2''
(figure~\ref{fig:6V2}). 


\section{CHMM approach underpinnings in the Baum-Welch algorithm}
\label{app:B}

We recall some basic features of the Baum-Welch (forward-backward) algorithm and show how they affect the constrained version of the HMMs. 

Let us consider a sequence of hidden states $(z_1, \ldots, z_T)$, where each $z_t\in\{Z_1, Z_2, \ldots, Z_N\}$, and a sequence of observed states $(x_1, \ldots, x_T)$ (for us, $x_t$ is the recorded experimental intensity). We assume that the initial state is chosen with a probability $\vec{\pi} = (\pi_1, \dots, \pi_N)$. Furthermore, given state $z_t = i$, we assume that the observable $x_t$ follows the {\em emission} distribution $x_t \sim b_i(x) = p(x|Z_i)$ (for us, $b_i(x) \sim \exp[-(x-\mu_i)^2/2\sigma_i^2]$, with $\mu_i$ the mean of the emitted signal from the hidden state $Z_i$, and $\sigma_i^2$ its variance).

For the forward phase, we define $\alpha_t(i) \equiv p(x_1, \ldots, x_t, z_t = i)$. The value of $\alpha_1(i)$ is computed as $\alpha_1(i) = \pi_i b(x_1)$. The subsequent values of $\alpha_t(i)$ are then recursively evaluated as
\begin{equation}
\alpha_t(i) =\sum_{j= 1}^N \alpha_{t-1}(j)a_{ji}b_i(x_t)
\end{equation}
In the backward phase, we use another auxiliary variable $\beta_t(i) = p(x_{t+1}, \dots, x_T | z_t = i)$. Also $\beta$'s are evaluated recursively: starting from the last value $\beta_T(i) = 1$, one iteratively computes the previous ones as 
\begin{equation}
\beta_t(i) = \sum_{j = 1}^{N} a_{ij}b_j(x_{t+1})\beta_{t+1}(j)
\end{equation}
The state-posterior $\gamma_t(i) = p(q_t = i | x_1, \dots, x_T)$ is then evaluated as 
\begin{equation}\gamma_t(i) = \frac{\alpha_t(i)\beta_t(i)}{\sum_{j= 1}^N \alpha_t(j)\beta_t(j)}\end{equation}
The transition rates are then updated according to the rule
\begin{equation}
    \centering
    a_{ij} = \frac{\sum_{t=1}^{T-1}\xi_{t}(i,j)}{\sum_{t=1}^{T-1}\gamma_t(i)}
    \label{eq: transition_matrix_update}
\end{equation}
where $\xi_{t}(i,j)$ is evaluated as 
\begin{equation}
    \centering
    \xi_{t}(i,j) = a_{ij} \frac{\alpha_t(i)\beta_{t+1}(j)b_j(y_{t+1})}{\sum_k \sum_w \alpha_t(k)a_{kw}\beta_{t+1}(w)b_w(y_{t+1})}
    \,.
\end{equation}
In a CHMM, for some pair $ij$ of states, we may have that
\begin{equation}
a_{ij} = 0 \Longleftrightarrow \xi_{t}(i,j) = 0, \text{ } \forall t.
\end{equation}
This, together with equation \eqref{eq: transition_matrix_update}, implies that if $a_{ij}$ is initialized as zero to avoid $i\to j$ transitions in the CHMM, it will remain zero at every update. 

\vspace{4mm}


\providecommand{\newblock}{}

\end{document}